\documentclass[twocolumn,showpacs,preprintnumbers,amsmath,amssymb]{revtex4}

\usepackage{graphicx}
\usepackage{dcolumn}
\usepackage{bm}

\begin{document}

\title{Generation of pulse trains by current-controlled magnetic mirrors}

\author{A. A. Serga}
\email{serga@physik.uni-kl.de}
\author{T. Neumann}
\author{A. V. Chumak}
\author{B. Hillebrands}
\affiliation{%
Fachbereich Physik and Forschungszentrum OPTIMAS\\
Technische Universit\"at Kaiserslautern, 67663 Kaiserslautern, Germany}%

\date{\today}

\begin{abstract}
The evolution of a spin-wave packet trapped between two direct current-carrying wires placed on the
surface of a ferrite film is observed by Brillouin light scattering. The wires act as semi-transparent
mirrors confining the packet. Because the spin-wave energy partially passes through these mirrors,
trains of spin-wave packets are generated outside the trap. A numerical model of this process is
presented and applied to the case when the current in the wires is dynamically controlled. This
dynamical control of the mirror reflectivity provides new functionalities interesting for the field of
spin-wave logic like that of a spin-wave memory cell.
\end{abstract}

\pacs{75.30.Ds, 85.70.Ge}
\maketitle

Spin-wave logic has attracted much attention, especially since the experimental realization of XOR- and
NAND-gates \cite{Kos05, Vas07, Sch08, Lee08}. The principle design of these basic elements relies on the
interaction of a spin-wave packet propagating in a waveguide with the magnetic inhomogeneity formed by
the Oersted field of a direct current-carrying wire placed on the waveguide surface. This interaction
allows one to control the spin-wave amplitude and phase as necessary \cite{Dem04,Kos07}.

For several reasons a main focus of interest is given to multi-wired structures, which represent a
particular realization of so called magnonic crystals \cite{Pus03}. Such structures potentially decrease
the required current. Moreover, since they can trap spin-wave packets, they possess new functionalities,
for instance, as spin-wave memory elements, spin-wave delay lines or multipliers.

Here, we investigate the propagation of a spin-wave packet across a double-wire structure consisting of
two parallel, current-conducting wires placed on the surface of a spin-wave waveguide. Different from
the observation of resonant tunneling \cite{Han07} when a static current was applied to both wires and,
consequently, the spin-wave packet had to pass both of them simultaneously, we dynamically switch the
current in one wire in such a way, that the spin-wave packet is trapped between the wires. In the
experiment the generation of pulse trains in transmission and reflection was observed by time- and
space-resolved Brillouin light scattering spectroscopy. Numerical simulations on the basis of our
previous experimental data \cite{Neu09} obtained for a single-wire structure support the results. In
particular, the time interval between two consecutively generated packets is mainly determined by the
distance between the wires while their relative intensity is governed by the applied direct currents. As
a consequence, by dynamically changing the applied direct currents the intensity of the spin-wave
packets can be modified according to any desired functional dependence. This is illustrated by
discussing the generation of pulse trains consisting of spatially or temporally equal pulses.

\begin{figure}[t]
\includegraphics[height = 30ex]{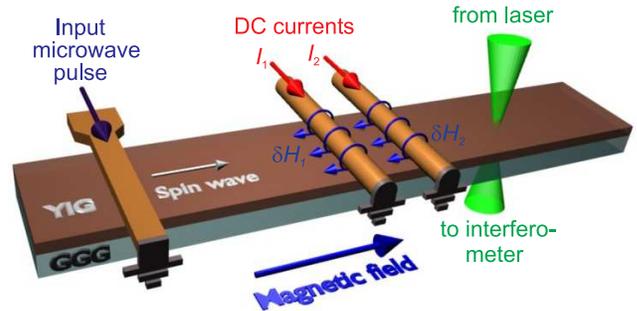}
\caption{\label{fig:aufbau} Sketch of the experimental section.}
\end{figure}

Figure~1 shows the experimental setup using a macroscopic yttrium iron garnet (YIG) stripe for easy
experimental conditions. An $18~{\rm ns}$ long input microwave pulse with a carrier frequency
$7.125~{\rm GHz}$ was sent to an input microstrip antenna placed on the surface of an YIG waveguide. The
$1.5~{\rm mm}$ wide and $25~{\rm mm}$ long YIG-film of thickness $5.7~\mu{\rm m}$ was magnetized along
its longitudinal axis by a magnetic field $H=1836~{\rm Oe}$. In the chosen configuration the microwave
pulse excites a packet of backward volume magnetostatic spin waves with a wave vector $k = 112~{\rm rad
/ cm}$ parallel to the bias magnetic field and a group velocity of $30~\mu{\rm m / ns}$.

In the center of the YIG-film, two parallel $50~\mu{\rm m}$ thick wires were placed at a distance of
$0.8~{\rm mm}$ from each other. To these wires direct currents $I_1 = 1.2~{\rm A}$ and $I_2 = 0.8~{\rm
A}$ were applied. While current $I_2$ was constantly operating, current $I_1$ was dynamically switched
on only after the spin-wave pulse had passed the wire. This way the spin-wave pulse was trapped between
two barriers formed by the current-carrying wires. We emphasize at this point that the polarity of the
currents was chosen in such a way, that the magnetic field locally decreases the bias field and a
barrier for the spin-wave propagation is formed \cite{Dem04}. This ensures that the transmission of the
spin-wave packets depends monotonically on the applied current and that the frequency and, consequently,
the wave-number dependence of the transmission is negligible \cite{Neu09}. The latter aspect is
noteworthy considering the short pulse duration which was used.

\begin{figure}[t]
\includegraphics[height = 72ex]{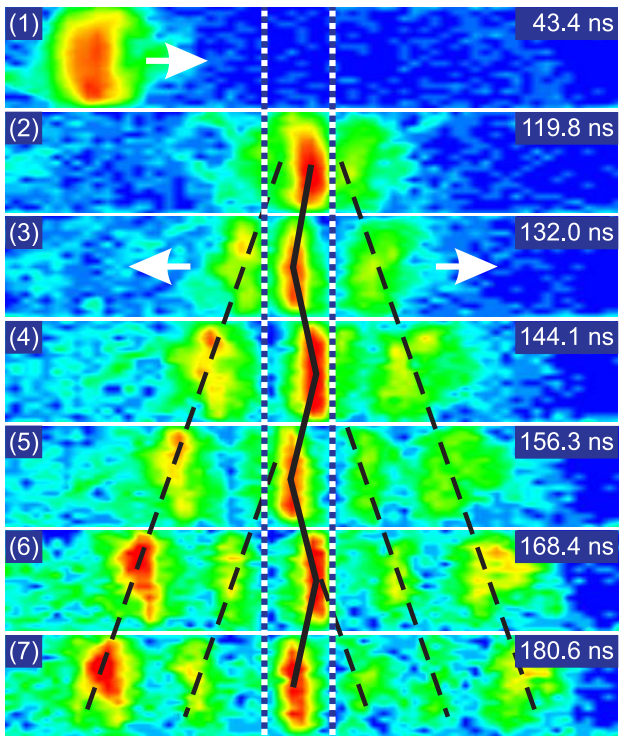}
\caption{\label{fig:BLSresults} (Color online) Space-resolved Brillouin light scattering images of the
$6~{\rm mm}$ wide central part of the sample at different times after the initial pulse was launched.
The dotted vertical lines indicate the positions of the wires. The dashed diagonal lines and the central
zigzag line illustrate the evolution of the generated wave packets and of the initial pulse,
respectively.}
\end{figure}

The time and spatial evolution of the spin-waves packet was detected by means of time- and
space-resolved Brillouin light scattering \cite{Bue00}. The obtained results are shown in Fig.~2. Each
frame represents the space distribution of the dynamic magnetization in the sample detected at a
specific moment in time as indicated in the right upper corner. The initial pulse (Frame~1) propagates
from left to right and passes the first wire - to which at this time no current is applied. The pulse is
reflected from the barrier formed by the second current-carrying wire (Frames~2 and 3). At this point,
current $I_1$ is switched on. Consecutively, the spin-wave packet starts wobbling between the two
current-carrying wires which is illustrated by the central zigzag line in Fig.~2.

Each time the pulse is reflected from one of the two wires, a certain fraction of the spin-wave energy
tunnels through the barrier (see for instance Frame~4). The transmitted part of the spin-wave packet is
detected as a pulse of the same duration as the original pulse outside of the trap. It continues its
propagation to either side of the sample as indicated by the dashed lines. Overall, the generation of
five, clearly distinguishable spin-wave packets was observed. These five packets as well as the central
trapped spin-wave pulse are shown in Fig.~3(a), which contains a horizontal cut obtained by summing up
the intensities in Frame~5 along a vertical line. Each packet is identified by a number, which indicates
the position in the sequence, at which the packet was generated.

In order to model the pulse train generation, we write the initial amplitude $A(x,t=0)$ of the wave
packet as
\begin{displaymath}
A(x, t=0) = \sum_{k} A_{k} e^{ik x}
\end{displaymath}
where the coefficients $A_{k}$ are found by a Fourier transformation of the rectangular excitation
pulse. Damping is included in this formulation as an imaginary contribution to the spin-wave frequency
$\omega$.

The amplitude distribution after a specific time $t$ is then given as the sum of partial amplitudes from
different transmitted and reflected waves. Hence, for the central section of the sample between the two
wires we have $A(x,t) = A_{+}(x,t) + A_{-}(x,t)$ with two contributions, one summand from waves
propagating in the same direction as the initial pulse
\begin{eqnarray*}
A_{+}(x,t) &=& \sum_{k} \big(A_{k} + A_{k} R_k^{(1)} R_k^{(2)} e^{i 2kx_0} \\
                &+& A_{k} (R_k^{(1)})^2  (R_k^{(2)})^2  e^{i 4kx_0} + \ldots \big)
                 e^{i(\omega(k) t + k x)}\\
&=& \sum_{k} A_{k} \frac{1}{1 - R_k^{(1)}  R_k^{(2)}  e^{i 2kx_0}} e^{i(\omega(k) t + k x)}
\end{eqnarray*}
and one summand from waves reflected on the second wire and propagating in the opposite direction
\begin{eqnarray*}
A_{-}(x,t) = \sum_{k} A_{k} \frac{R_k^{(1)} e^{i2kx_0}}{1 - R_k^{(1)}  R_k^{(2)}  e^{i 2kx_0}}
e^{i(\omega(k) t - k x)}.
\end{eqnarray*}
Here, we have chosen $x = 0$ for simplicity to coincide with the position of the first wire. $x_0$
denotes the distance between the two wires, which restricts the region of applicability for the last two
equations to $0 < x < x_0$. In the equations, $R_k^{(1)}$ and $R_k^{(2)}$ are factors describing the
reflection of the spin-wave amplitude for the first and second wire, respectively.

Accordingly, equations can be found which involve the amplitude transmission coefficients $T_k^{(1,2)} =
\sqrt{1 - R_k^{(1,2)^2}}$ in order to describe the spin-wave amplitude to the right of the second wire
(i.e. for $x > x_0$) and to the left of the left of the first wire ($x < 0$).

\begin{figure}[t]
\includegraphics[height = 70ex]{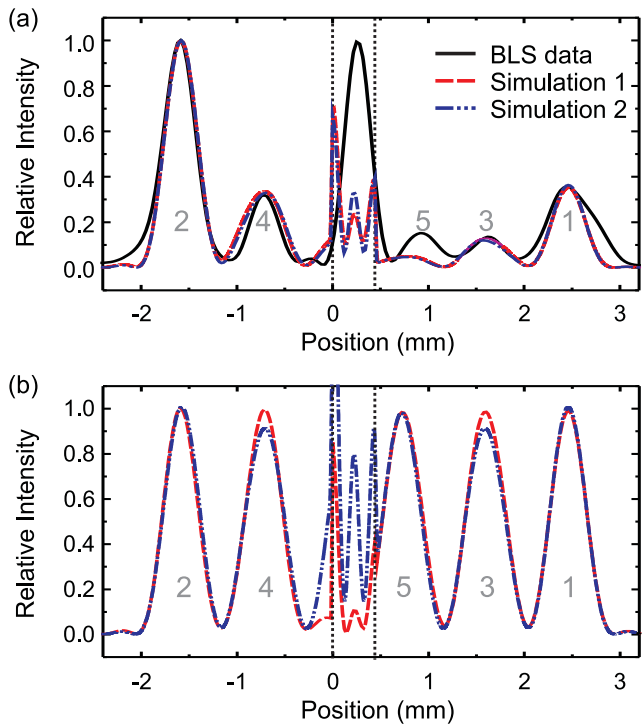}
\caption{\label{fig:simulation} (Color online) (a) Comparison of the intensity profile from Brillouin
light scattering measurements after $180.6~{\rm ns}$ with numerical simulations. (b) Numerical
simulations with dynamically adjusted currents and reflection coefficients.}
\end{figure}

In the simulation, the signal pulse duration was set to $18~{\rm ns}$ and the distance between the wires
was chosen as $x_0 = 0.44~{\rm mm}$. This value is smaller than the geometric distance between the wires
because the implicit assumption made above, that the reflection takes place at the position of the wire,
is only approximately satisfied. In fact, the larger the currents, the more the current-induced Oersted
fields of the wires reduce the area of localization where the spin-wave packet is trapped.

Two types of simulations were performed. Simulation~1 used previously measured data for a single
current-carrying wire \cite{Neu09} from which the reflection coefficients $R_k^{(1,2)}$ were
interpolated. In order to determine the influence of the $k$-dependence of $R$ on the results, a second
set of simulations was run using $k$-independent coefficients $R_k^{(1,2)} = R^{(1,2)}$ (Simulation~2).
The values of $R$ used in these simulations can be considered as averaged reflection coefficients and
display the ratio of energy transmitted and reflected at the wires more clearly than the current values
do.

The results of these simulations are presented in Fig.~3(a). There is a large discrepancy for the
central peak which is a result of the simplicity of the proposed model.

The intensity of the central peak is very sensitive to the exact interference conditions for the spin
waves reflected from the wires. However, the phase change due to the current induced barrier is not
included in our considerations, neither is any change in wave-vector or a resulting change in group
velocity caused by the magnetic inhomogeneity. Despite these limitations, the simulations for the
generated pulses agree well with the experimental data which shows that the model is actually working.

For suitable parameters $R^{(1)} = 0.69~{\rm A}$ and  $R^{(2)} = 0.92~{\rm A}$, Simulation~1 and
Simulation~2 in Fig.~3 can be brought to perfect coincidence. From this, we conclude that it is indeed
possible to replace the $k$-dependent sets $R_k^{(1,2)}$ by two $k$-independent parameters. We would
like to attract the reader's attention to the fact, that the reflection on wire~2 is higher than on
wire~1 - which results in the first generated spin-wave packet being smaller than the second one -
though the current applied to wire~2 is smaller. The reason for this effect is the thermal heating of
the sample around the wire~2 due to the constantly applied current which leads to an increased
scattering.

One main advantage of the section design is the controllability of the current on very short time scales
compared to the spin-wave relaxation time. As a consequence, the current-carrying wires can be used like
a set of mirrors, whose reflectivity can be adjusted dynamically. Due to the high reflectivity, that can
be reached, it is possible to store a signal packet for a certain period of time between the mirrors and
afterwards release it. This realizes the functionality of a spin-wave memory cell. Of course, the
storage time of the spin-wave packet is limited by the damping of the packet inside the trap. However,
it is possible to compensate the spin-wave damping between the magnetic mirrors, for instance by means
of parametric amplification \cite{Mel99}.

Even without parametric amplification it is possible to use the dynamic properties of the mirrors in
order to gain functionality. As a first example, we consider the generation of pulse trains consisting
of spatially separated pulses with equal intensity for a fixed time.

Figure~3(b) shows two simulations of such pulse trains, which were generated by a time-dependent
variation of the reflection characteristics. As previously, the parameters $R_k^{(1,2)}$ in Simulation~1
are taken from experimental data. The current was manually adjusted to generate pulses of equal height.
In Simulation~2 the $k$-independent coefficients $R^{(1,2)}$ were chosen according to

\begin{equation}\label{Eq:Spatial}
R(1) = R \quad \mbox{and} \quad R(n+1) = \sqrt{1 - \frac{1-R^2}{1 - n(1-R^2)}}
\end{equation}
where $n$ is the number of the packet which is generated. This analytically guarantees the same
amplitudes for all wave numbers and all generated pulses. In Simulation~2, $R = 0.92$ was chosen so that
the first pulse has the same intensity as the pulse 1 in Fig.~3(a).

Clearly, the number of spin-wave packets with an equal intensity, which can be generated by simply
turning down the current and, thereby, adjusting the reflectivity is limited. According to
Eq.~(\ref{Eq:Spatial}), the maximal number $n_{max}$ is a function of the initial reflection coefficient
$R$, which determines the intensity of the first packet
\begin{displaymath}
n_{max} = \lfloor \frac{R^2}{1-R^2} \rfloor +1.
\end{displaymath}

For microwave applications it is more interesting to require that the temporally separated pulses picked
up at an antenna (which is located at a fixed position in space) are of equal intensity. In this case,
in Eq.~(\ref{Eq:Spatial}) $R(n+1)$ has to be modified to
\begin{displaymath}
\sqrt{1 - \frac{(1-R^2)e^{\frac{2n \Gamma x_0}{v_{\rm G}}}}{1 - (1 + e^{\frac{2 \Gamma x_0}{v_{\rm G}}}+
\ldots + e^{\frac{2 (n-1)\Gamma x_0}{v_{\rm G}}})(1-R^2)}}
\end{displaymath}
where $\Gamma$ is the relaxation frequency and $v_{\rm G}$ the group velocity of the spin-wave packet.

In conclusion, we have experimentally demonstrated the generation of pulse trains from a single
spin-wave packet by using two current-carrying wires as magnetic mirrors between which the spin-wave
packet is trapped.

The performed numerical simulations based on the presented model are in good agreement with the
experiment. It was shown, that for the tunneling regime, when the current creates a barrier for the
spin-wave propagation, the frequency dependence of the transmission and reflection can be neglected and
is replaced by a single parameter even for short pulses with a wide Fourier spectrum.

By dynamically changing the current, the generated pulse trains can be modified in any desired way. This
was illustrated by simulations with variable reflection coefficients. In particular, the important
examples of pulse trains consisting of pulses with equal intensity which are observed (i) at different
positions in space but at a fixed moment in time and (ii) at different times but at a fixed position in
space were discussed.

Our work has been financially supported by the Matcor Graduate School of Excellence and DFG SE 1771/1-1.

%
%
%
%
%

\end{document}